\documentstyle[preprint,aps]{revtex}
\begin{document}
\tightenlines
\title{New gauge conditions in general relativity: what can we
learn from them?}
\author{Giampiero Esposito \thanks{Electronic address:
giampiero.esposito@na.infn.it} and Cosimo Stornaiolo
\thanks{Electronic address: cosmo@na.infn.it}}
\address{Istituto Nazionale di Fisica Nucleare, Sezione di Napoli,
Mostra d'Oltremare Padiglione 20, 80125 Napoli, Italy\\
Dipartimento di Scienze Fisiche, Universit\`a degli Studi di
Napoli Federico II, Complesso Universitario di Monte S. Angelo,
Via Cintia, Edificio G, 80126 Napoli, Italy}
\maketitle

\begin{abstract}
The construction of conformally invariant gauge conditions for
Maxwell and Einstein theories on a manifold $M$ 
is found to involve two basic 
ingredients. First, covariant derivatives of a linear 
gauge (e.g. Lorenz or de Donder), completely contracted with
the tensor field representing the metric on the vector bundle of
the theory. Second, the addition of a compensating term, obtained
by covariant differentiation of a suitable tensor field built from
the geometric data of the problem. If the manifold $M$
is endowed with a $m$-dimensional 
positive-definite metric $g$, the existence
theorem for such a gauge in gravitational theory can be proved. 
If the metric $g$ is Lorentzian, which corresponds to general
relativity, some technical steps are harder, but one has again to
solve integral equations on curved space-time to be able to impose
such gauges.
\end{abstract}

Our recent work on gauge conditions in (linearized) general relativity
has been motivated by the Eastwood--Singer [1] derivation of
conformally invariant gauge conditions for vacuum Maxwell theory. We
are now going to describe the key steps of our construction.

For vacuum Maxwell theory in four dimensions in the absence of
sources, the operator acting on the potential $A_{b}$ is well 
known to be
\begin{equation}
P_{a}^{\; b}=-\delta_{a}^{\; b}\Box
+R_{a}^{\; b}+\nabla_{a}\nabla^{b},
\label{(1)}
\end{equation}
where $\nabla$ is the Levi--Civita connection on space-time,
$\Box \equiv g^{ab} \nabla_{a} \nabla_{b}$, and $R_{a}^{\; b}$
is the Ricci tensor. Thus, the supplementary (or gauge) condition
of the Lorenz type, i.e.
\begin{equation}
\nabla^{b}A_{b}=0,
\label{(2)}
\end{equation}
is of crucial importance to obtain a wave equation for $A_{b}$.
The drawback of Eq. (2), however, is that it is not preserved
under conformal rescalings of the metric:
\begin{equation}
{\widehat g}_{ab}=\Omega^{2} g_{ab}, \; \; \; \; 
{\widehat g}^{ab}=\Omega^{-2} g^{ab},
\label{(3)}
\end{equation}
whereas the Maxwell equations
\begin{equation}
\nabla^{b}F_{ab}=0
\label{(4)}
\end{equation}
are invariant under the rescalings (3). This remark was the
starting point of the investigation 
by Eastwood and Singer [1], who
found that a conformally invariant supplementary condition
may be imposed, i.e.
\begin{equation}
\nabla_{b} \left[\Bigr(\nabla^{b}\nabla^{a}-2R^{ba}
+{2\over 3}R g^{ba} \Bigr)A_{a} \right]=0.
\label{(5)}
\end{equation}
As is clear from Eq. (5), conformal invariance is achieved at the
price of introducing third-order derivatives of the potential.
In flat backgrounds, such a condition reduces to
\begin{equation}
\Box\nabla^{b}A_{b}=0.
\label{(6)}
\end{equation}
Of course, all solutions of the Lorenz gauge are also solutions
of Eq. (6), whereas the converse does not hold.

Leaving aside the severe technical problems resulting from the attempt 
of quantizing in the Eastwood--Singer gauge [2], we are now interested
in understanding the key features of the counterpart for Einstein's
theory of general relativity. In other words, although the vacuum
Einstein equations
\begin{equation}
R_{ab}-{1\over 2}g_{ab}R=0
\label{(7)}
\end{equation}
are not invariant under the conformal rescalings (3), we would like
to see whether the geometric structures 
leading to Eq. (5) admit
a non-trivial generalization to Einstein's theory, so that a
conformally invariant supplementary condition with a higher order
operator may be found as well. For this purpose, we re-express 
Eqs. (2) and (5) in the form
\begin{equation}
g^{ab}\nabla_{a}A_{b}=0,
\label{(8)}
\end{equation}
\begin{equation}
g^{ab}\nabla_{a}\nabla_{b}\nabla^{c}A_{c} 
+ \nabla_{b} \left[\Bigr(-2R^{ba}+{2\over 3}R g^{ba}\Bigr)A_{a}
\right]=0.
\label{(9)}
\end{equation}
Equation (8) involves the space-time metric in its contravariant
form, which is also the metric on the bundle of 1-forms on $M$.
In Einstein's theory, one deals instead with the vector bundle of 
symmetric rank-2 tensors on space-time
with DeWitt supermetric
\begin{equation}
E^{abcd} \equiv {1\over 2} \Bigr(g^{ac}g^{bd}
+g^{ad}g^{bc}+ \alpha g^{ab}g^{cd} \Bigr),
\label{(10)}
\end{equation}
$\alpha$ being a real parameter different from $-{2\over m}$,
where $m$ is the dimension of space-time (this restriction on
$\alpha$ is necessary to make sure that the metric $E^{abcd}$
has an inverse). One is thus led to replace
Eq. (8) with the de Donder gauge
\begin{equation}
W^{a} \equiv E^{abcd}\nabla_{b}h_{cd}=0.
\label{(11)}
\end{equation}
Hereafter, $h_{ab}$ denotes metric perturbations
in linearized general relativity.
The supplementary condition (11) is not invariant under conformal
rescalings, but the expression of the Eastwood--Singer gauge in
the form (9) suggests considering as a ``candidate'' for a
conformally invariant gauge involving a higher-order operator
the equation
\begin{equation}
E^{abcd}\nabla_{a}\nabla_{b}\nabla_{c}\nabla_{d}W^{e}
+ \left[\Bigr(\nabla_{p}T^{pebc}\Bigr)+T^{pebc}\nabla_{p}
\right]h_{bc}=0 .
\label{(12)}
\end{equation}
More precisely, Eq. (12) is obtained from Eq. (9) by applying
the replacement prescriptions 
$$
g^{ab} \rightarrow E^{abcd},
$$
$$
A_{b} \rightarrow h_{ab}, 
$$
$$
\nabla^{b}A_{b} \rightarrow W^{e},
$$ 
with $T^{pebc}$ a rank-4 tensor field obtained from
the Riemann tensor, the Ricci tensor, the trace of Ricci and
the metric. In other words, $T^{pebc}$ is expected to involve
all possible contributions of the kind $R^{pebc}, R^{pe} g^{bc},
R g^{pe} g^{bc}$, assuming that it should be linear in 
the curvature. The analysis of the full theory, however,
shows that $T^{pebc}$ is even more involved.

Indeed, if $\gamma$ is a metric solving the full Einstein
equations in vacuum, and $g$ is a background metric, a gauge
condition linear in $\gamma$ which reduces to Eq. (12) in the
linearized approximation may be written in the form [3]
\begin{equation}
S^{e}(\gamma) \equiv E^{abcd}(g)\nabla_{a}\nabla_{b}
\nabla_{c}\nabla_{d}W^{e}(\gamma) 
+ \nabla_{p}{\widetilde T}^{pe}(\gamma)=0,
\label{(13)}
\end{equation}
where
\begin{equation}
W^{e}(\gamma) \equiv E^{epqr}(g)\nabla_{p}\gamma_{qr},
\label{(14)}
\end{equation}
and the connection $\nabla$ annihilates $g$ but not $\gamma$.
${\widetilde T}^{pe}(\gamma)$ is a rank-2 tensor field to be
determined (see below). We now study the behaviour of 
$S^{e}(\gamma)$ under conformal rescalings of the physical metric 
$\gamma$, since it is $\gamma$ which solves the vacuum Einstein
equations. It is then convenient to denote by $Q^{e}(\gamma)$ the
first term on the right-hand side of Eq. (13). On the one hand,
the invariance of $S^{e}$ under conformal rescalings of $\gamma$
means that
\begin{equation}
S^{e}(\Omega^{2}\gamma)=\Omega^{2}\Bigr(Q^{e}(\gamma)
+\nabla_{p}{\widetilde T}^{pe}(\gamma)\Bigr).
\label{(15)}
\end{equation}
On the other hand, the explicit calculation shows that
\begin{equation}
S^{e}(\Omega^{2}\gamma)=\Omega^{2}Q^{e}(\gamma)+U^{e}
+\nabla_{p}{\widetilde T}^{pe}(\Omega^{2}\gamma),
\label{(16)}
\end{equation}
where $U^{e}$ depends on $\Omega, \gamma$ and their covariant
derivatives up to the fourth order. Its lengthy expression can be
found in Sec. 5 of [3]. By virtue of (15) and (16), and of the
identity
\begin{equation}
\nabla_{p}\Bigr(\Omega^{2}{\widetilde T}^{pe}(\Omega^{2}\gamma)\Bigr)
= 2\Omega \Omega_{;p}{\widetilde T}^{pe}(\gamma) 
+ \Omega^{2}\nabla_{p}{\widetilde T}^{pe}(\gamma),
\label{(17)}
\end{equation}
the desired equation for ${\widetilde T}^{pe}(\gamma)$ reads
\begin{equation}
\nabla_{p}\Bigr[{\widetilde T}^{pe}(\Omega^{2}\gamma)
-\Omega^{2}{\widetilde T}^{pe}(\gamma)\Bigr] 
+ 2\Omega \Omega_{;p} {\widetilde T}^{pe}(\gamma) 
= -U^{e}.
\label{(18)}
\end{equation}
Since ${\widetilde T}^{pe}(\gamma)$ can be arbitrarily chosen, it
is sufficient to show that a {\it particular} class of such tensors
exists for which Eq. (18) (and hence Eq. (15)) is satisfied. For
this purpose, we assume that
\begin{equation}
\nabla_{q}\Bigr[{\widetilde T}^{pe}(\Omega^{2}\gamma)
-\Omega^{2}{\widetilde T}^{pe}(\gamma)\Bigr] 
+ 2\Omega \Omega_{;q} {\widetilde T}^{pe}(\gamma) 
= -{1\over m}\delta_{q}^{\; p} U^{e}.
\label{(19)}
\end{equation}
Both sides of Eq. (19) can be contracted with a vector $f^{q}$
on $(M,g)$. After defining the operator
\begin{equation}
{\cal D} \equiv f^{q}\nabla_{q},
\label{(20)}
\end{equation}
together with (the symbol $\otimes_{s}$ represents symmetrized 
tensor product)
\begin{equation}
{\widetilde T}_{(\;)} \equiv {\widetilde T}_{pe}dx^{p}
\otimes_{s} dx^{e}={\widetilde T}_{(pe)} dx^{p}
\otimes_{s} dx^{e},
\label{(21)}
\end{equation}
\begin{equation}
{\widetilde T}_{\wedge} \equiv {\widetilde T}_{pe}
dx^{p} \wedge dx^{e}={\widetilde T}_{[pe]}dx^{p}\wedge dx^{e},
\label{(22)}
\end{equation}
\begin{equation}
(fU)_{(\; )} \equiv f_{p}U_{e}dx^{p} \otimes_{s} dx^{e}
=f_{(p} \; U_{e)} dx^{p} \otimes_{s}dx^{e},
\label{(23)}
\end{equation}
\begin{equation}
(fU)_{\wedge} \equiv f_{p}U_{e}dx^{p}\wedge dx^{e}
=f_{[p} \; U_{e]} dx^{p} \wedge dx^{e},
\label{(24)}
\end{equation}
and introducing the kernel (with our notation, $G_{\cal D}(x,y)$ is
the Green kernel of the operator $\cal D$)
\begin{equation}
K_{\Omega}(x,y) \equiv 
2 G_{\cal D}(x,y)[\Omega{\cal D}\Omega](y) 
- \delta(x,y)\Omega^{2}(y),
\label{(25)}
\end{equation}
we eventually derive from Eq. (18) the integral equation
\begin{equation}
[{\widetilde T}_{\diamond}(\Omega^{2}\gamma)](x)
+ \int_{M}K_{\Omega}(x,y)[{\widetilde T}_{\diamond}(\gamma)](y)dV(y)
= -{1\over m} \int_{M}G_{\cal D}(x,y)(fU)_{\diamond}(y)
dV(y),
\label{(26)}
\end{equation}
where the symbol $\diamond$ is a concise notation for the subscript
$(\;)$ or $\wedge$ used in (21)--(24). The right-hand side of 
Eq. (26) is completely known for a given choice of the vector $f^{p}$
and of the dimension $m$ of $M$.

If the metric $\gamma$ is positive-definite, and if the operator
$\cal D$ defined in (20) is symmetric and elliptic on a compact 
Riemannian manifold $M$ without boundary, with $f^{p}$ so chosen
that $\cal D$ has no zero-modes, the solution of Eq. (26) can be
reduced to the task of solving an infinite system of algebraic
equations (see Eq. (5.40) of [3]).

In our analysis of conformal invariance of gauge conditions, it is
crucial to consider conformal rescalings of the physical metric
$\gamma_{ab}$, while the background metric $g_{ab}$ is kept fixed.
We have done so because it is $\gamma_{ab}$ which solves the 
Einstein equations, which are not conformally invariant. The
consideration of general mathematical structures seems to suggest
that a key ingredient is the addition of a ``compensating term'' 
$\nabla_{p}{\widetilde T}^{pe}(\gamma)$ to the higher-order
covariant derivatives of the original gauge condition (see Eqs. (9)
and (13)). Unlike the case of Maxwell theory in curved backgrounds,
where conformal rescalings of the background metric are considered,
we have therefore studied conformal rescalings of the physical
metric only in general relativity. Still, it remains of interest
for further research to consider conformal rescalings of both 
background and physical metric. 

It also remains to be seen how to extend our results [3], which
hold for positive-definite metrics, to the case of Lorentzian
metrics, which are of course the object of interest in general
relativity. We can however point out that the integral equation
resulting from Eq. (18) does not depend on the signature of the 
metric, and hence the construction of 
${\widetilde T}^{pe}(\gamma)$ remains non-local also in the
Lorentzian case [3]. The Green kernels that one may want to use
will be distinguished by various boundary conditions, and hence
the Lorentzian framework will be actually richer in this respect.

It therefore seems that new perspectives are in sight in the
investigation of gauge conditions in general relativity. They
might be applied both in classical theory (linearized equations 
in gravitational wave theory, symmetry principles), and in the
attempts of quantizing the gravitational field [3].

\acknowledgements
This work has been partially supported by PRIN97 ``Sintesi''.

\end{document}